\documentclass[runningheads]{llncs}
\usepackage[T1]{fontenc}
\usepackage{graphicx}
\usepackage{booktabs}
\usepackage[misc]{ifsym}
\usepackage{amsmath}
\usepackage{amssymb}
\usepackage[utf8]{inputenc}
\usepackage{booktabs}
\usepackage{amssymb} 
\usepackage{hyperref}
\usepackage{multirow}
\usepackage{makecell}
\newcommand{\corr}{(\Letter)}
\usepackage{mwe}

\begin{document}

\title{PolyBench: Benchmarking LLM Forecasting and Trading Capabilities on Live Prediction Market Data}

\titlerunning{PolyBench: LLM Forecasting and Trading on Prediction Markets}

\author{Pu Cheng\inst{1}, Juncheng Liu\inst{1}, Yunshen Long\inst{2} \corr}

\authorrunning{P. Cheng et al.}

\institute{College of Electrical Engineering, Sichuan University, Chengdu, Sichuan, China 
\email{\{cheng.chengpu123,junchengliu2004\}@gmail.com} \and
School of Economics, Sichuan University, Chengdu, Sichuan, China 
\email{dhhlys@163.com}}

\maketitle              

\begin{abstract}
Predicting real-world events from live market signals demands systems that fuse qualitative news with quantitative order-book dynamics under strict temporal discipline---a challenge existing benchmarks fail to capture.
We present \textbf{PolyBench}, a multimodal benchmark derived from Polymarket that records point-in-time cross-sections of 38,666 binary prediction markets spanning 4,997 events, synchronously coupling each snapshot
with a Central Limit Order Book~(CLOB) state and a real-time news stream.
Using PolyBench, we evaluate seven state-of-the-art Large Language Models---spanning open- and closed-source families---generating 36,165 predictions under identical, timestamp-locked market states collected between February~6 and~12,~2026.
Our multidimensional framework assesses directional accuracy, our proposed Confidence-Weighted Return~(CWR), Annualized Percentage Yield~(APY), and Sharpe ratio via realistic order-book execution simulation.
The results reveal a pronounced performance divergence: only two of seven models achieve positive financial returns---MiMo-V2-Flash at \textbf{17.6\%} CWR and Gemini-3-Flash at 6.2\% CWR---while the remaining five incur losses despite uniformly high stated confidence.
These findings highlight the gap between surface-level language fluency and genuine probabilistic reasoning under live market uncertainty, and establish PolyBench as a contamination-proof, financially-grounded evaluation standard for future LLM research.
Our dataset and code available at \underline{\href{https://github.com/PolyBench/PolyBench}{https://github.com/PolyBench/PolyBench}}.

\end{abstract}

\section{Introduction}

Forecasting the future is a longstanding scientific challenge, ranging from early astronomy to modern econometric modeling.
In machine learning, it encompasses time-series analysis
\cite{box1976time,lim2021time}, online learning \cite{foster1999regret}, and conformal prediction \cite{barber2023conformal}.
Yet \emph{open-domain} forecasting---producing reliable probabilistic predictions across diverse, unseen topics without domain-specific fine-tuning---remains largely unsolved.
Achieving such foresight at scale would represent a paradigm shift in artificial intelligence, with significant implications for market efficiency, risk management, and data-driven policy \cite{arrow2008promise}.

Large Language Models~(LLMs) are natural candidates for this task.
Trained on vast corpora via next-token prediction, they exhibit emergent reasoning capabilities that extend well beyond surface-level pattern matching \cite{bubeck2023sparks}, suggesting that predicting the next token may generalize toward predicting the next real-world \emph{event}.
Recent studies \cite{zou2022forecasting,halawi2024approaching} have
demonstrated encouraging progress, yet a central question remains open:
\textit{Can AI systems accurately forecast outcomes and execute profitable trades on live decentralized prediction markets using multimodal data?}

Decentralized prediction markets such as Polymarket
\cite{saguillo2025unravelling} constitute a uniquely rigorous evaluation environment---financially incentivized, on-chain, and structurally immune
to answer memorization.
Forecasting in this setting is inherently multimodal: participants must jointly assess qualitative resolution criteria, breaking news, and quantitative Central Limit Order Book~(CLOB) states, bid-ask spreads, and live liquidity.
Existing benchmarks reduce forecasting to static text question answering or numerical time-series extrapolation, thereby overlooking the order-book mechanics, capital-risk dimension, and point-in-time grounding that real prediction markets require.

To address this gap, we present a dual-purpose research initiative:

\begin{figure}[h!]
  \centering
  \includegraphics[width=\linewidth]{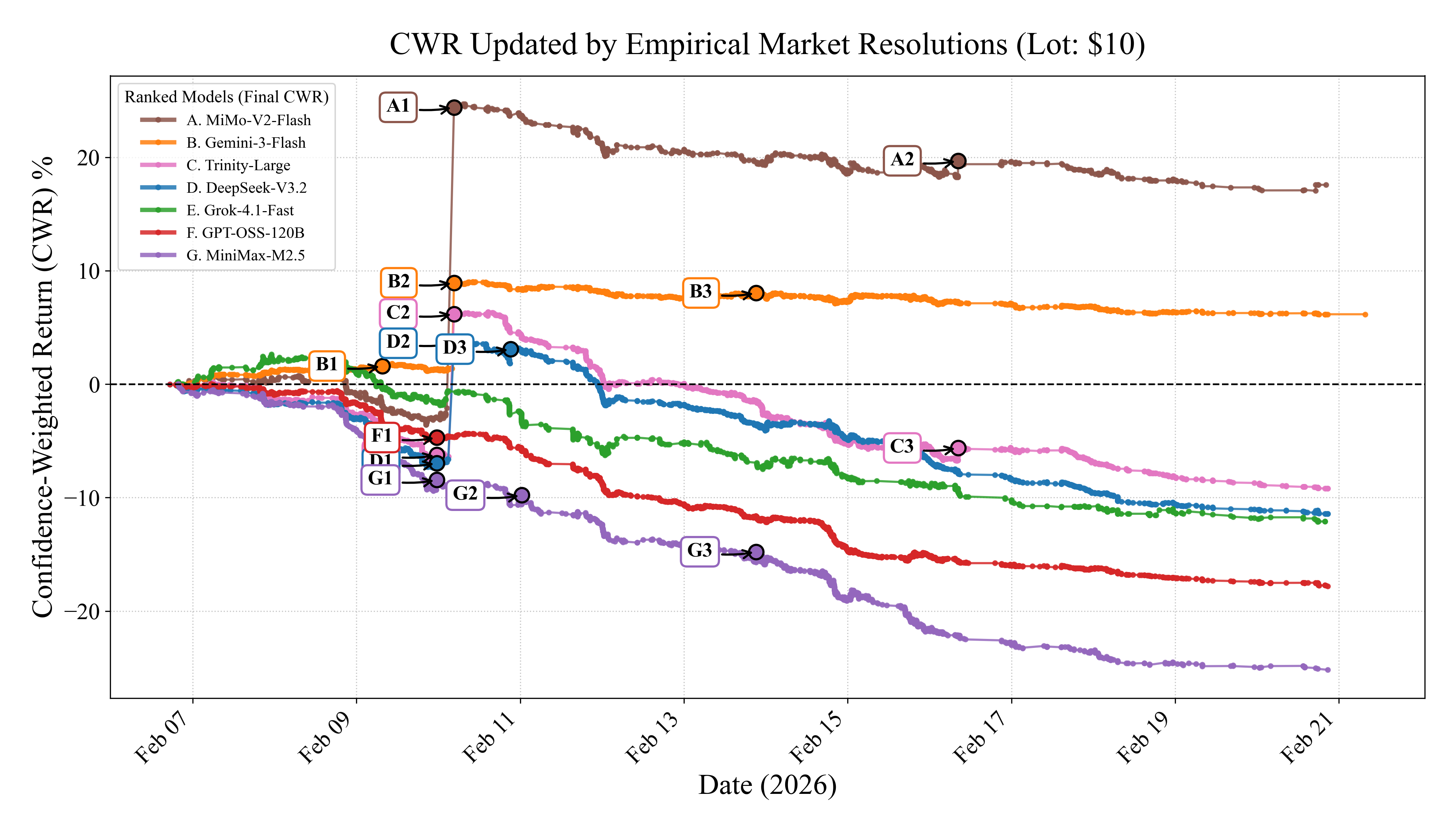}
  \caption{Confidence-Weighted Return (CWR) timeline reflecting empirical portfolio evolution. Annotated markers denote outsized individual trade returns resulting from correct, high-confidence predictions on low-probability events. Only \textit{MiMo-V2-Flash} and \textit{Gemini-3-Flash} sustain positive trajectories, isolating predictive alpha from market noise.}
  \label{fig:returns_over_time}
\end{figure}

\begin{itemize}
    \item \textbf{PolyBench}---a live multimodal benchmark comprising 38,666 binary prediction markets across 4,997 real-world events. Each market snapshot
    synchronously couples an exact CLOB state with contemporaneous news and official resolution criteria.
    \item \textbf{Multidimensional Evaluation Framework}---a protocol that assesses seven state-of-the-art LLMs on both forecasting accuracy and trading viability, measuring Confidence-Weighted Return (CWR), Annualized
    Percentage Yield (APY), Sharpe ratio, and instruction adherence under realistic order-book execution simulation.
\end{itemize}

To quantify LLM viability in financial forecasting, our multidimensional framework assesses classification accuracy, probabilistic reliability, strategic reasoning, and simulated financial execution, highlighting both emerging strengths and current limitations in information aggregation.
As shown in Fig. \ref{fig:returns_over_time}, our results reveal a pronounced performance divergence: only two of the seven evaluated models achieve positive financial returns, while the remaining five incur losses. demonstrating that strong language modeling does not guarantee profitable trading under live market conditions.

\section{Related Work}
\label{sec:related}

Deploying LLMs on live decentralized prediction markets lies at the
intersection of three active research areas: financial time-series
modeling, contamination-free LLM evaluation, and financial agent
benchmarking.

\subsection{Event Forecasting and Financial Time-Series}

Computational finance has evolved from classical econometric models
\cite{bollerslev1986generalized,box1970time} through recurrent and
attention-based deep learning \cite{hochreiter1997long,vaswani2017attention}
to large-scale financial foundation models such as BloombergGPT
\cite{wu2023bloomberggpt}, FinBERT \cite{araci2019finbert},
and Time-LLM \cite{jin2023time}.
However, these works predominantly target \emph{continuous} asset
variables rather than discrete event outcomes.
NLP-oriented forecasting benchmarks such as Autocast
\cite{zou2022forecasting} address binary event prediction but lack
real-time quantitative market grounding.
PolyBench bridges this gap by coupling synchronous order-book snapshots
with aligned news streams to enable evaluation of binary event outcomes
under realistic financial conditions.

\subsection{Overcoming Pre-training Data Leakage}

Static benchmarks are vulnerable to data contamination: models may
inadvertently memorize evaluation instances during pre-training, even
in carefully curated test sets such as MATH~\cite{hendrycks2021measuringmathematicalproblemsolving}
or Humanity's Last Exam~\cite{phan2025humanitylastexam}.
Dong et al.\ \cite{dong2024generalizationmemorizationdatacontamination} show that
this leakage can inflate reported accuracies by up to 66.9\%, rendering
published benchmark scores unreliable proxies for true generalization.
Decentralized prediction markets are structurally immune to this problem,
because they forecast events that have \emph{not yet occurred} at
evaluation time, making it impossible to pre-train on the ground-truth
outcomes.
PolyBench therefore enforces genuine zero-shot temporal reasoning and
provides a contamination-proof measure of true LLM forecasting capability.

\subsection{LLMs as Autonomous Forecasting Agents}

Recent work has shifted toward evaluating LLMs as autonomous
decision-making agents \cite{liu2023agentbench}.
Halawi et al.\ \cite{halawi2024approaching} showed that LLM-based systems
can approach human-crowd-level accuracy on Polymarket;
Prophet Arena \cite{yang2025llmasaprophet} further exposes predictive
bottlenecks on live events.
However, both works evaluate models on purely textual axes, ignoring
order-book mechanics and the financial cost of trade execution.
PolyBench addresses this limitation by treating prediction markets as
an inherently multimodal environment---ingesting CLOB spreads and live
liquidity alongside breaking news---and assessing models on the metric
that ultimately matters in practice: simulated financial return.

Table~\ref{tab:benchmarks} summarizes the feature gap that PolyBench fills.

\begin{table}[htbp]
\centering
\caption{Comparison of prediction-market and forecasting benchmarks.
 PolyBench is the only benchmark combining Polymarket data, live order books,
 aligned news, and return-based evaluation at scale.}
\label{tab:benchmarks}

\resizebox{\textwidth}{!}{
\begin{tabular}{l @{\hspace{.5cm}} c @{\hspace{.5cm}}  c @{\hspace{.5cm}}  c @{\hspace{.5cm}} c @{\hspace{.5cm}} c}
\toprule
Method & Markets & \makecell{w/ \\ Polymarket} & \makecell{w/ \\ Order Book} & \makecell{w/ \\ News} & \makecell{CWR \\ Eval.} \\
\midrule
Time-LLM \cite{jin2023time} & -- & $\times$ & $\times$ & $\times$ & $\times$ \\
AgentBench \cite{liu2024deepseek} & -- & $\times$ & $\times$ & $\times$ & $\times$ \\
FinBERT \cite{araci2019finbert} & -- & $\times$ & $\times$ & \checkmark & $\times$ \\
BloombergGPT \cite{wu2023bloomberggpt} & -- & $\times$ & $\times$ & \checkmark & $\times$ \\
Autocast \cite{zou2022forecasting} & 6{,}707 & $\times$ & $\times$ & \checkmark & $\times$ \\
Prophet Arena \cite{yang2025llmasaprophet} & 1{,}367 & $\times$ & $\times$ & \checkmark & $\times$ \\
Halawi et al.\ \cite{halawi2024approaching} & 48{,}754 & \checkmark & $\times$ & \checkmark & $\times$ \\
LiveTradeBench \cite{yu2025livetradebench} & 10 & \checkmark & \checkmark & \checkmark & $\times$ \\
\textbf{PolyBench (Ours)} & \textbf{38{,}666} & \checkmark & \checkmark & \checkmark & \checkmark \\
\bottomrule
\end{tabular}
}
\end{table}

\section{Polymarket Definitions and Notations}

To formalize the evaluation framework within PolyBench, we define the core components of the Polymarket ecosystem. While these definitions share foundational concepts with general prediction markets, they are specifically adapted to the decentralized mechanisms, order book structures, and snapshot-based evaluations utilized in our pipeline.

\noindent\textbf{Event.} Let $\{E_i\}_{i=1}^K$ denote the set of evaluated forecasting events sourced via Polymarket's Gamma API. An event serves as the overarching high-level container concerning a future real-world occurrence (e.g., political elections, financial milestones, or sports outcomes). It establishes the context, scope, and strict official resolution criteria for the markets beneath it. 
\begin{itemize}
    \item \textit{Example:} ``Who will win the 2026 World Cup?''
\end{itemize}

\noindent\textbf{Market.} Each event $E_i$ contains one or more binary markets $\{M_{ij}\}_{j=1}^{N_i}$. A market is a specific, actively tradable proposition under an event that will ultimately resolve to a definitive Yes (True) or No (False). Markets are identified by unique token identifiers and are accompanied by live volume metrics and Central Limit Order Book (CLOB) states. 
\begin{itemize}
    \item \textit{Example market:} ``France will win the 2026 World Cup.''
\end{itemize}

\noindent\textbf{Event Resolution.} An event $E_i$ is said to resolve at time $\tau_i$ when all of its constituent markets $M_{ij}$ have their outcomes $o_{ij} \in \{0,1\}$ definitively settled according to Polymarket's official resolution rules. An outcome of $o_{ij} = 1$ indicates that the market $M_{ij}$ resolved to Yes. PolyBench explicitly accounts for strict adherence to these rules, including conditional constraints (e.g., forced 50-50 split resolutions).

\noindent\textbf{Contract (Share).} For a given market $M_{ij}$, a Yes contract (or share) is a binary asset that yields a payout of 1 if $o_{ij} = 1$, and 0 otherwise. Rather than occupying a static theoretical price, the acquisition cost of these shares is governed dynamically by available market liquidity at a specific point in time. In PolyBench, the executed share price is simulated directly against the historical Central Limit Order Book (CLOB), sequentially absorbing orders from the best available ask ($a_{ij}$) until the agent's allocated capital limit is reached.

\noindent\textbf{State Snapshot.} Because PolyBench evaluates trading agents in a simulated historical context, each market $M_{ij}$ is evaluated within a temporally anchored State Snapshot. This snapshot is a multi-modal bundle containing the exact historical timestamp, event metadata, precise CLOB state (including wide spreads or volume imbalances), and conditionally prefetched exogenous Google News context. This ensures a strict Bayesian baseline and prevents future data leakage during backtesting.

\noindent\textbf{Trading Signal.} The actionable output generated by the LLM agent for a given snapshot, characterized by a scalar confidence score $c_i \in [0.0, 1.0]$. The agent issues a BUY decision only when $c_i$ overcome a rigid threshold 0.6.

\section{PolyBench}

The goal of PolyBench is to provide a return-focused, extendable and automated evaluation of LLMs' advanced forecasting and market trading capabilities on Polymarket, aiming to match or even surpass the expertise of human professionals. In this section, we will introduce the details of PolyBench.

\begin{figure}[t]
  \centering
  \includegraphics[width=\linewidth]{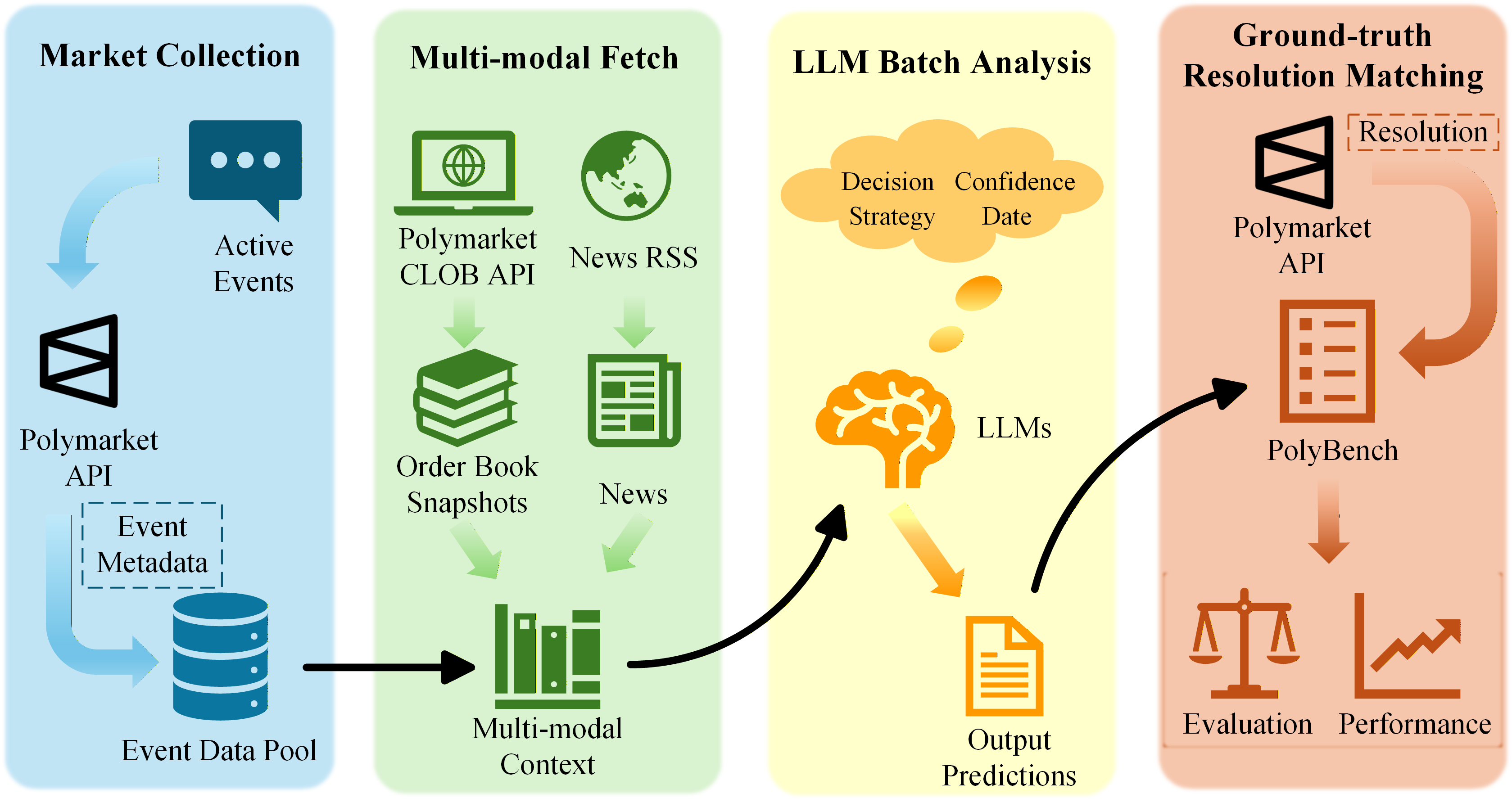}
  \caption{The four-stage PolyBench construction pipeline:
           (1)~market collection via the Polymarket Gamma API,
           (2)~multi-modal fetch of news and order-book snapshots,
           (3)~LLM batch analysis, and
           (4)~ground-truth resolution matching.}
  \label{fig:pipeline}
\end{figure}

\subsection{Construction}

As illustrated in Fig. \ref{fig:pipeline}, the benchmark construction pipeline initiates with the collection of active events via Polymarket's Gamma API, extracting essential metadata such as titles, token identifiers, and market volumes. A background runner then aggregates multi-modal context by retrieving relevant Google News article streams and capturing precise Central Limit Order Book (CLOB) states. This temporally aligned bundle of information constitutes a historical \textit{snapshot} of each market at a specific timestamp. These multi-modal state snapshots are subsequently dispatched to the respective LLM endpoints for predictive evaluation.

\subsubsection{Prompt Design}
To accurately simulate algorithmic trading without future data leakage during backtesting, the prompt establishes a strict Bayesian baseline locked to the specific snapshot timestamp. The language model is anchored with the exact historical timestamp as the ``Current Date'', alongside the event title and detailed description. The prompt injects the historical market options—including precise order book spreads and midpoint prices active at that exact interval—coupled with the conditionally prefetched exogenous news context. 

To govern the agent's decision-making process, a robust system instruction enforces the following core rules:
\begin{itemize}
    \item \textbf{Resolution Rule Primacy}: Strict adherence to official market rules over general knowledge, explicitly accounting for conditional constraints (e.g., forced 50-50 split resolutions).
    \item \textbf{Value Identification}: Directives to signal ``value bets'' only when the model's evidence-backed posterior significantly deviates from the market's implied odds.
    \item \textbf{Market Microstructure Analysis}: Guidelines to interpret order book data, recognizing wide spreads as uncertainty and volume imbalances as informed trading.
    \item \textbf{Evidence-Backed Conviction}: A hard constraint requiring probability confidence exceeding 60\% ($c > 0.6$) to be directly justified by provided news quotes.
    \item \textbf{Temporal Awareness}: Continuous evaluation of the estimated resolution timeline against the ``Current Date'' snapshot for annualized yield mapping.
\end{itemize}

\subsubsection{Structured Output and Filtering}
The LLM predictive engine must return a strictly formatted JSON payload with an overall reasoning narrative and an array of actionable \texttt{decisions}. Each executable trade within this array must adhere to the schema detailed in Table \ref{tab:output_schema}.

\begin{table}[h!]
\centering
\caption{Structured JSON schema requirements for model predictions.}
\begin{tabular}{lp{7cm}}
\hline
{Field} & {Description} \\ \hline
\texttt{decision} & The specific target option identifier \\ 
\texttt{outcome} & The exact predicted position mapping \\ 
\texttt{strategy} & A classification of the trading rationale (e.g., \texttt{value\_bet}, \texttt{news\_catalyst}) \\
\texttt{confidence} & A scalar conviction score ($c \in [0.0, 1.0]$) \\
\texttt{est\_resolution\_date} & The model's estimated temporal timeline for market settlement \\
\texttt{reasoning} & A concise, specific rationale validating the position \\ \hline
\end{tabular}

\label{tab:output_schema}
\end{table}

Crucially, if a model fails to identify any positive expected value trades or if its declared confidence falls below a rigid threshold ($c < 0.6$), the decisions are procedurally discarded. A model can suggest skipping a event explicitly by giving nothing in decision field. This functionally mimics a disciplined trader's behaviour, abstaining from ambiguous markets without penalty.

\subsubsection{Dataset Statistics}
The executed data collection pipeline results in an expansive evaluation framework. Table \ref{tab:dataset_overview} provides a comprehensive breakdown of the curated PolyBench dataset, highlighting the scale of real-world financial events, gathered snapshots, and the resulting density of large language model predictions processed. Notably, all events, underlying prediction markets, and corresponding snapshots---including both exogenous news context and internal order book states---were collected between February 6, 2026, and February 12, 2026. The latest market resolution date recorded is February 21, 2026. Crucially, the inclusion of actively traded, yet-to-be-resolved markets imbues PolyBench with inherent extensibility for future longitudinal assessments.

\begin{table}[t]
  \centering
  \caption{PolyBench dataset statistics.}
  \label{tab:dataset_overview}
  \begin{tabular}{lrr}
    \toprule
    {Component} & {Count} & {Notes} \\
    \midrule
    Events              & 4,997   & \\
    \quad Active        & 2,904   & Open at collection time \\
    \quad Resolved      & 2,093   & Closed / expired \\
    \midrule
    Markets             & 38,666  & \\
    \quad Active        & 26,784  & Open at collection time \\
    \quad Inactive      & 11,882  & Settled or delisted \\
    \midrule
    Market Snapshots    & 38,666  &\\
    \quad With news context   & 35,436 & 91.6\% coverage \\
    \quad With order book     & 19,012 & 49.2\% coverage \\
    \quad Prediction-ready    & 5,097  & Resolved + With order book \& news \\
    \midrule
    LLM Predictions     & 36,165  & $\sim$5,100 per model \\
    \bottomrule
  \end{tabular}
\end{table}

\subsection{Evaluation}

We continue to use model accuracy as an indicator of forecasting capabilities,
in line with recent LLM benchmark studies \cite{zeng2025futurexadvancedlivebenchmark,karger2025forecast}.
These studies typically require models to predict a deterministic outcome for evaluation. However, a fundamental limitation of this approach is that forecasting inherently differs from standard classification tasks due to the probabilistic and random nature of future events \cite{yang2025llmasaprophet}.
Furthermore, given the lack of reliable, unbiased continuous probability references in volatile decentralized prediction markets, we do not adopt traditional proper scoring rules such as the Brier Score. Instead, we directly incorporate the model's stated confidence into our final economic return calculation to better evaluate the financial viability of each trade suggestion.

Unlike traditional forecasting-oriented benchmarks that force models to predict every event, PolyBench is fundamentally tailored toward realistic market trading and maximizing practical economic returns. Consequently, agents are provided with a \texttt{SKIP} choice to explicitly pass on ambiguous or unfavorable markets, and these instances are entirely excluded from the performance metrics calculation.

\subsubsection{Directional Accuracy}

To assess a model's baseline forecasting capabilities on its actively chosen positions, we rely on the standard metric of directional accuracy. For a set of $N$ valid, non-skipped predictions, accuracy is defined as the proportion of trades where the model correctly identifies the ultimate resolution of the prediction market. Let $o_k$ represent the ground-truth terminal settlement state for the $k$-th forecasted market, and $\hat{o}_k$ represent the agent's forecasted position.

\begin{equation}
    \text{Accuracy} = \frac{1}{N} \sum_{k=1}^{N} \mathbb{I}(o_k = \hat{o}_k)
\end{equation}

Where $\mathbb{I}(\cdot)$ is the indicator function that evaluates to 1 if the predicted state matches the actual settlement, and 0 otherwise. While accuracy serves as a fundamental performance baseline, it treats all correct guesses equally and fails to distinguish between predictions made with marginal certainty and those made with absolute conviction.

\subsubsection{Confidence-Weighted Return (CWR)}
To capture both correctness and risk calibration under realistic market constraints, we introduce the Confidence-Weighted Return (CWR) metric. Rather than evaluating a flat percentage return against mid-market prices, the agent's simulated capital allocation is scaled linearly by its stated \texttt{confidence} ($c_k \in [0.0, 1.0]$) and a designated base \texttt{lot\_size} ($L$) for each \texttt{BUY} decision. 

For the $k$-th predicted trade, the allocated budget is $B_k = c_k \times L$. This budget is deployed against the corresponding active Central Limit Order Book (CLOB) snapshot, absorbing available liquidity starting from the best ask. This simulates a realistic sweep of the order book, returning the actual invested capital $I_k \le B_k$ and the total number of shares acquired $S_k$. 

If the model accurately predicts the market's resolution, the shares settle at exactly \$1.00, yielding a gross payout. If incorrect, the shares resolve at \$0.00. The net profit $\pi_k$ for trade $k$ is thus:

\begin{equation}
    \pi_k = 
    \begin{cases} 
        S_k - I_k, & \text{if prediction is correct} \\
        -I_k,     & \text{if prediction is incorrect}
    \end{cases}
\end{equation}

Formally, the aggregate CWR across $N_{buy}$ executed trades is defined as the total net profit divided by the total invested capital:

\begin{equation}
    \text{CWR} = \frac{\sum_{k=1}^{N_{buy}} \pi_k}{\sum_{k=1}^{N_{buy}} I_k}
\end{equation}

By structurally accounting for variable slippage and execution limits, this metric systematically rewards models that exhibit strong risk calibration---placing high confidence on successful bets with adequate liquidity---while severely penalizing overconfident allocations on incorrect assumptions.

\subsubsection{Additional Economic and Temporal Metrics}
To holistically evaluate the practical viability and risk-adjusted performance of the agents' trading strategies, we incorporate three supplementary continuous metrics:

\begin{itemize}
    \item \textbf{Annualized Percentage Yield (APY):} While CWR measures absolute return, APY normalizes this return over time to account for the opportunity cost of locked capital. For a trade yielding return $r_i$ held for $D$ days before its actual resolution, its annualized baseline is computed as $\text{APY} = r_i \times (365 / D)$.
    \item \textbf{Sharpe Ratio:} To assess risk-adjusted portfolio performance, we compute the Sharpe Ratio over the distribution of a model's individual trade returns. Defined as the mean return divided by the standard deviation of returns ($\mu_r / \sigma_r$), it strictly penalizes models experiencing erratic, high-variance equity draw-downs.
    \item \textbf{Temporal Resolution Error:} Beyond predicting outcomes, agents must forecast \textit{when} a market will resolve to gauge capital efficiency. This metric represents the mean absolute error (in days) between the model's \texttt{est\_\allowbreak resolution\_\allowbreak date} and the true market resolution timestamp.
\end{itemize}

\section{Experiment}
\label{sec:experiment}

In this section, we present the empirical results from the PolyBench evaluation framework. The benchmark evaluates seven distinct LLMs representing state-of-the-art architectures from their respective providers, evaluating a gross total of 36,165 predictions based on active financial snapshots.

Table \ref{tab:model_statistics} summarizes the evaluated models, including their releasing company, the specific API service provider utilized, reported quantization applied during inference, cost of predicting over 5097 prediction-ready markets and open-source states as shown in Table \ref{tab:dataset_overview}. To the best of our knowledge, all models' knowledge cut-off date is before the benchmark.

\begin{table}[h!]
\centering
\caption{Evaluated models grouped by open-source availability, detailing respective companies, API providers, quantization formats, and their costs.}
\label{tab:model_statistics}
\begin{tabular}{lllllc}
\toprule
{Model Name} & {Company} & {Service Provider} & {Quantization} & {Cost} & {Open Source}\\
\midrule
GPT-OSS-120B & OpenAI & DeepInfra & FP4 & \$ 1.34 & \multirow{5}{*}{Yes} \\
DeepSeek-V3.2 & DeepSeek & Atlas Cloud & FP8 & \$ 3.32 & \\
MiMo-V2-Flash & Xiaomi & Xiaomi & FP8 & \$ 1.72 & \\
MiniMax-M2.5 & MiniMax & Inceptron & FP8 & \$ 8.43 & \\
Trinity-Large-Preview & Arcee AI & Prime & - & Free & \\
\midrule
Gemini-3-Flash-Preview & Google & Vertex AI & - & \$ 11.50 & \multirow{2}{*}{No} \\
Grok-4.1-Fast & xAI & xAI & - & \$ 6.85 & \\
\bottomrule
\end{tabular}
\end{table}

\subsection{Overall Performance}

To rigorously assess the financial capability and predictive accuracy of each LLM, we employ the metrics derived in Section 4. Crucially, we distinguish between standard, unweighted portfolio returns (\textit{Non-CWR}) and proposed \textit{Confidence-Weighted Return (CWR)}. Table \ref{tab:overall_results} details these comprehensive results of metrics at base lot size $L=\$10$ across all executed non-skipped trades.

\begin{table}[ht]
\centering
\caption{Overall LLMs performance on PolyBench. (Base Lot Size=\$10)}
\label{tab:overall_results}
\resizebox{\columnwidth}{!}{%
\begin{tabular}{lcccccc}
\toprule
{Model} & {Accuracy $\downarrow$} & {Avg.\ Conf.} & {Non-CWR} & {CWR} & {APY} & {Sharpe} \\
\midrule
Gemini-3-Flash & 75.0\% & 0.855 & 4.1\% & \underline{6.2\%} & 1486.5\% & 0.01 \\
MiMo-V2-Flash & 62.1\% & 0.859 & 11.1\% & \textbf{17.6\%} & 4067.3\% & 0.02 \\
Grok-4.1-Fast & 59.1\% & 0.862 & -15.9\% & -12.1\% & -5820.4\% & -0.17 \\
GPT-OSS-120B & 58.0\% & 0.836 & -19.4\% & -17.8\% & -7075.8\% & -0.22 \\
DeepSeek-V3.2 & 53.5\% & 0.838 & -15.3\% & -11.4\% & -5571.8\% & -0.03 \\
Trinity-Large & 42.7\% & 0.871 & -12.4\% & -9.2\% & -4527.8\% & -0.02 \\
MiniMax-M2.5 & 42.6\% & 0.808 & -26.1\% & -25.2\% & -9537.2\% & -0.22 \\
\bottomrule
\end{tabular}%
}
\end{table}

The separation between Non-CWR and CWR fundamentally exposes a model's risk calibration capabilities. Across the board, models consistently achieve a \textit{higher} CWR than their unweighted Non-CWR baseline, demonstrating inherent meta-cognition: LLMs effectively assign higher confidence to correct predictions while discounting misjudgments. Thus in the following discussion, we will be consistent with CWR metric. Among the seven, only \textit{MiMo-V2-Flash} and \textit{Gemini-3-Flash} achieved positive absolute returns, yielding 17.6\% and 6.2\% CWR, respectively. Notably, the open-source \textit{MiMo-V2-Flash} demonstrated robust calibration dynamics, outperforming its proprietary counterparts in net profitability.

The trajectory of these returns is further contextualized in Figure \ref{fig:returns_over_time}, plotting CWR from the initial snapshot batch on February 6 through the progressive settlement of markets on February 21, 2026. Each point on the line stands for the settlement of one market. Annotated spikes denote outsized returns (exceeding 1,000\%) from successful low-probability predictions.

\subsubsection{Impact of Position Sizing and Order Book Slippage}

A critical dynamic introduced by PolyBench's realistic execution environment is the impact of position sizing on achievable returns. Unlike theoretically unlimited flat-price assumptions, trading against historical CLOB states accurately models liquidity constraints and execution slippage. To reliably model this dynamic without exceeding the top five levels of order-book depth captured in our snapshot pipeline, we observe simulated returns up to a maximum base lot size of \$1,000.

\begin{figure}[h!]
  \centering
  \includegraphics[width=\linewidth]{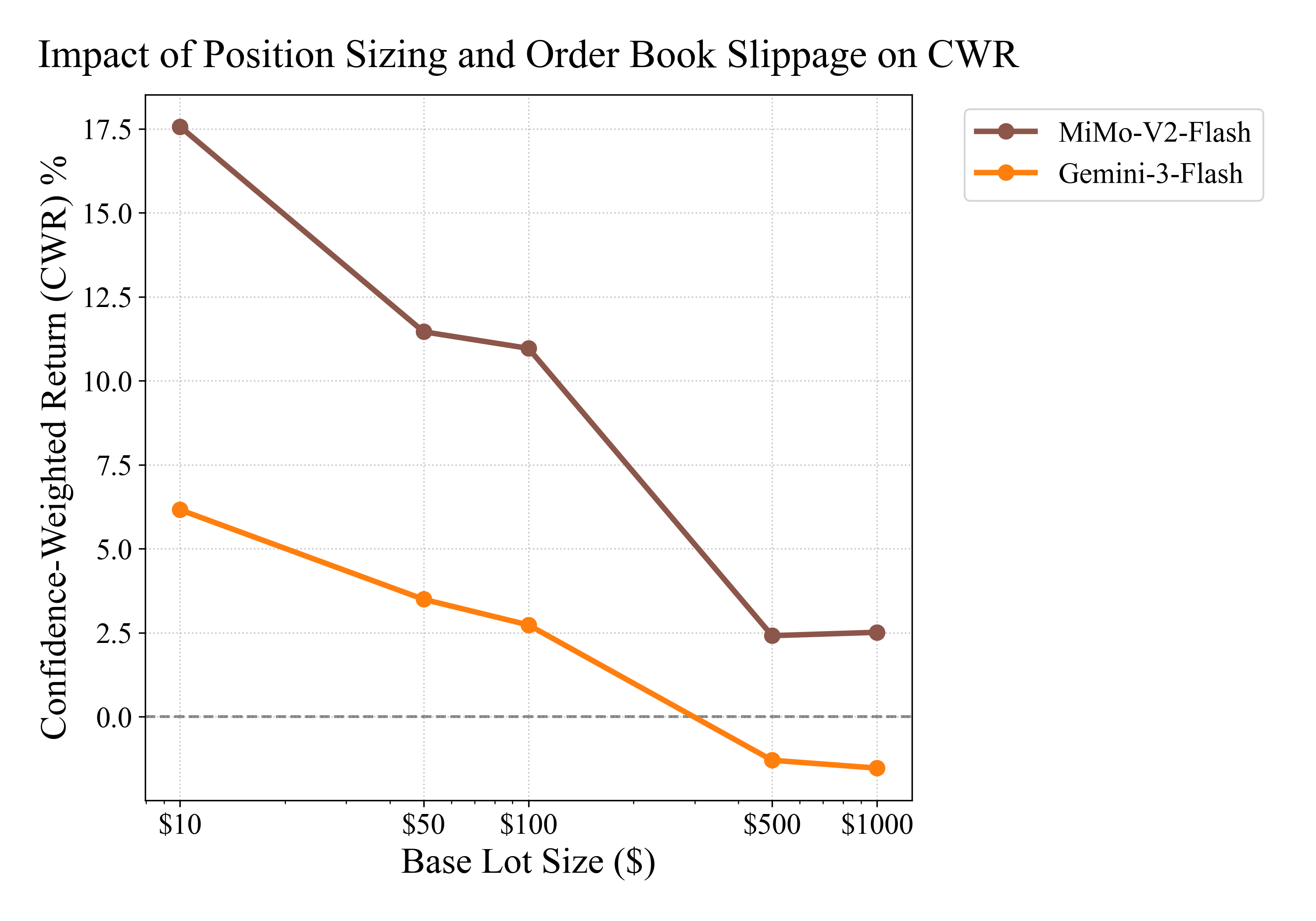}
  \caption{Impact of base lot size ($L$) on Confidence-Weighted Return (CWR) for \textit{MiMo-V2-Flash} and \textit{Gemini-3-Flash}. As the investment budget scales from \$10 to \$1,000, algorithmic execution slippage against the limited top levels of the historical order book rapidly decays theoretical alpha.}
  \label{fig:cwr_vs_lot}
\end{figure}

Figure \ref{fig:cwr_vs_lot} illustrates the degradation of CWR across varying base lot sizes ($L$) for the two top-performing models. At smaller capital allocations ($L \le \$100$), the limited required volume allows agents to absorb the immediate best-ask prices, preserving their theoretical alpha. However, as position sizing scales logarithmically toward \$1,000, model profitability violently contracts. High-volume trades exhaust the top levels of the captured order book, forcing execution into wider spreads at significantly worse average execution prices. This structural slippage erodes the alpha of \textit{Gemini-3-Flash}, while \textit{MiMo-V2-Flash} demonstrates relative resilience up to $\$500$ before ultimately succumbing to systemic liquidity limits. These results emphasize that in decentralized prediction markets, theoretical predictive advantages are strictly bounded by available market depth.

\subsection{Domain Expertise and Miscalibrated Conviction}

\begin{figure}[h!]
  \centering
  \includegraphics[width=\linewidth]{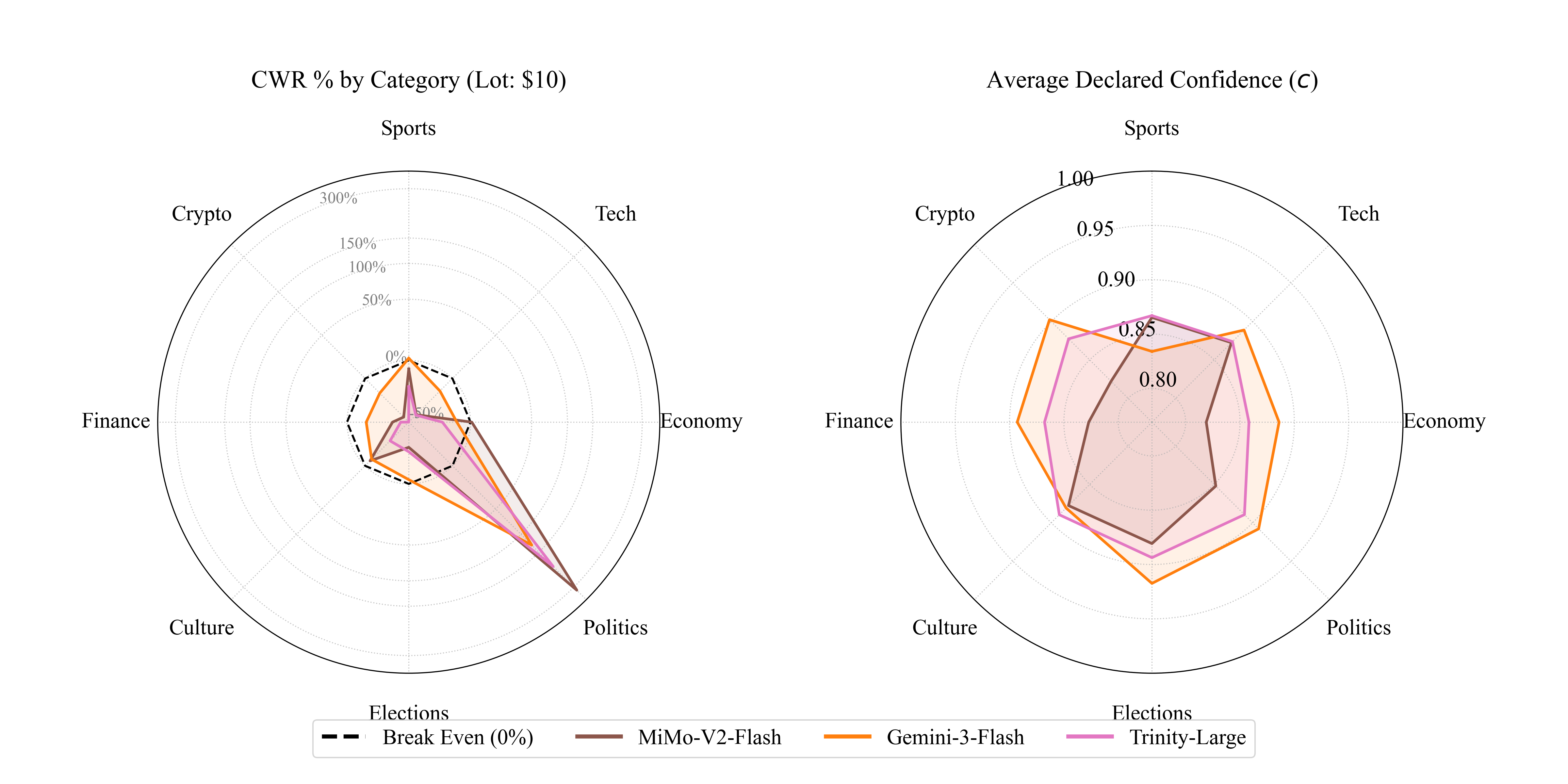}
  \caption{Dual-metric radar chart comparing empirical Confidence-Weighted Return (CWR) and Average Declared Confidence across eight event domains. The disparity highlights LLM miscalibration; models uniformly maintain high confidence ($c \ge 0.8$) across all domains, yielding severe negative returns in volatile sectors such as \textit{Crypto}.}
  \label{fig:radar_cwr_conf}
\end{figure}

Different thematic domains impart vastly different analytical challenges. Figure \ref{fig:radar_cwr_conf} maps both the CWR performance and the Average Declared Confidence of three top-performing models (\textit{MiMo-V2-Flash}, \textit{Gemini-3-Flash} and \textit{Trinity-Large}) against eight major event categories structuring PolyBench.

Figure \ref{fig:radar_cwr_conf} contrasts CWR performance and Average Declared Confidence across major event categories. The results expose a severe miscalibration: models maintain rigid, high confidence ($0.8 \le c \le 0.9$) regardless of domain complexity. While this conviction yields robust alpha in \textit{Politics}, a domain rich in parsable textual polling data, it proves highly detrimental in speculative, high-variance sectors such as \textit{Crypto}. Here, models incur deep negative returns while maintaining perilous overconfidence, underscoring that historical text is an insufficient proxy for volatile price-action derivatives.

\subsection{Strategy Utilization and Instruction Governance}

\begin{table}[h!]
\centering
\caption{Strategy utilization (\%) paired with subsequent financial performance (CWR\%, Sharpe Ratio), alongside execution instruction adherence (Invalid Strat.) in overall accuracy order.}
\label{tab:strategy_merged}
\resizebox{\linewidth}{!}{%
\begin{tabular}{l cccc c}
\toprule
{Model} & {Value Bet} & {News Cat.} & {Arbitrage} & {Stable Yld.} & {Invalid Strat.} \\
\midrule
Gemini-3-Flash & 80.2\% (\underline{7.2\%}, 0.01) & 40.0\% (\underline{20.7\%}, 0.03) & 0.8\% (\textbf{8.1\%}, 0.07) & 28.5\% (\underline{-0.1\%}, 0.00) & 8 \\
MiMo-V2-Flash & 94.8\% (\textbf{19.2\%}, 0.02) & 46.7\% (\textbf{50.0\%}, 0.04) & 5.7\% (-31.1\%, -0.37) & 0.7\% (\textbf{0.4\%}, 1.03) & - \\
Grok-4.1-Fast & 99.3\% (-11.7\%, -0.17) & 37.1\% (6.7\%, 0.04) & 3.5\% (-10.8\%, -0.11) & 0.2\% (-49.9\%, -0.70) & - \\
GPT-OSS-120B & 99.9\% (-17.8\%, -0.22) & 39.4\% (-19.3\%, -0.21) & 1.1\% (\underline{-2.7\%}, -0.05) & 2.5\% (-0.2\%, -0.02) & - \\
DeepSeek-V3.2 & 98.4\% (-10.9\%, -0.03) & 34.0\% (19.0\%, 0.02) & 15.3\% (-35.4\%, -0.37) & 20.7\% (-22.6\%, -0.29) & - \\ 
Trinity-Large & 98.2\% (-9.7\%, -0.03) & 38.1\% (16.6\%, 0.01) & 5.4\% (-10.3\%, -0.10) & 0.6\% (-38.5\%, -0.71) & 103 \\
MiniMax-M2.5 & 98.0\% (-25.0\%, -0.22) & 19.3\% (-19.8\%, -0.22) & 5.2\% (-22.7\%, -0.22) & 0.2\% (-40.0\%, -0.38) & 2 \\
\bottomrule
\end{tabular}%
}
\end{table}

Beyond evaluating pure predictive accuracy, a central tenet of autonomous financial deployment is rigorous instruction adherence. Table \ref{tab:strategy_merged} highlights strategy utilization, conditional performance, and execution discipline. LLMs overwhelmingly defaulted to \texttt{value\_bet} and \texttt{news\_catalyst}. However, \textit{Gemini-3-Flash} and \textit{DeepSeek-V3.2} demonstrated greater conceptual diversity, deploying \texttt{stable\_yield} in over 20\% of trades. Performance conditioned on strategy execution exposes LLMs' strategic mastery: \textit{MiMo-V2-Flash} generated notable alpha utilizing \texttt{news\_catalyst} (50.0\% CWR) and \texttt{value\_bet} (19.2\% CWR), while \textit{Gemini-3-Flash} achieved resilient positive returns across a broader set of domains.

Crucially, instruction adherence alone does not guarantee profitability. While extreme hallucinations (e.g., \textit{Trinity-Large} generating 103 invalid tags) predictably correlate with severe negative returns, flawless execution is insufficient for generating alpha. Models like \textit{GPT-OSS-120B} and \textit{Grok-4.1-Fast} achieved perfect compliance yet suffered sub-zero returns. In contrast, the open-source \textit{MiMo-V2-Flash} uniquely synthesized strict zero-hallucination discipline with precise predictive capability, generating the highest overall CWR. This demonstrates that empirical financial viability requires both rigorous operational governance and robust probabilistic reasoning.

\section{Conclusion}

We presented PolyBench, the first large-scale, contamination-proof benchmark that evaluates LLMs as trading agents on live decentralized prediction markets.
By coupling 38,666 binary market snapshots with point-in-time CLOB state and aligned news streams, PolyBench provides a financially-grounded evaluation standard that static benchmarks cannot replicate.

Across seven state-of-the-art LLMs, only \textit{MiMo-V2-Flash} (17.6\% CWR) and \textit{Gemini-3-Flash} (6.2\% CWR) achieve positive returns, while the remaining five incur losses despite uniformly high confidence.
Three key findings emerge: (1) LLMs exhibit implicit meta-cognition,
consistently assigning higher confidence to correct predictions; (2)~domain competence is highly uneven---strong textual signals in \textit{Politics} yield alpha, while speculative sectors such as \textit{Crypto} expose severe miscalibration; and (3)~instruction adherence is necessary but not sufficient for profitability.
Order-book slippage further erodes returns at larger lot sizes, bounding theoretical predictive advantages by available market depth.

\begin{credits}
\subsubsection{\ackname} This study is funded by National Social Science Fund of China [grant number 23XJY003].

\subsubsection{\discintname}
The authors have no competing interests to declare that are relevant to
the content of this article.
\end{credits}
%
%
%
\bibliographystyle{splncs04}
\bibliography{mybiblio}

\end{document}